\documentclass[conference]{IEEEtran}
\IEEEoverridecommandlockouts
\usepackage{cite}
\usepackage{amsmath,amssymb,amsfonts}
\usepackage{subfigure}
\usepackage{algorithmic}
\usepackage{caption}
\usepackage{graphicx}
\usepackage{subfig}
\usepackage{float}
\usepackage{textcomp}
\usepackage{xcolor}

\def\BibTeX{{\rm B\kern-.05em{\sc i\kern-.025em b}\kern-.08em
    T\kern-.1667em\lower.7ex\hbox{E}\kern-.125emX}}
\begin{document}

\title{Knowledge Graph Driven UAV Cognitive Semantic Communication Systems for Efficient Object Detection\\
}
\author{Xi Song$^{ \S }$, Lu Yuan$^{ \S }$, Zhibo Qu$^{ \S }$, Fuhui Zhou$^{ \S }$, Qihui Wu$^{ \S }$, Tony Q. S. Quek$^{ \ddag }$, and Rose Qingyang Hu$^{\dagger}$ \\
$^{ \S }$Nanjing University of Aeronautics and Astronautics, China, \\
$^{ \ddag }$Singapore University of Technology and Design, Singapore, \\
$^{\dagger}$Utah State University, USA \\
Email: \emph{\{sx32510, yuanlu, zhiboqu@nuaa.edu.cn, zhoufuhui@ieee.org,} \\
\emph{wuqihui2014@sina.com, tonyquek@sutd.edu.sg, and rosehu@ieee.org\}}
\thanks{This work was supported by National Key R\&D Program of China under Grant 2023YFB2904500, the National Natural Science Foundation of China under Grant 62222107, Grant 62071223 and Zhejiang Lab Open Research Project under Grant K2022PD0AB09.}
}

\maketitle
\begin{abstract}
Unmanned aerial vehicles (UAVs) are widely used for object detection. However, the existing UAV-based object detection systems are subject to the serious challenge, namely, the finite computation, energy and communication resources, which limits the achievable detection performance. In order to overcome this challenge, a UAV cognitive semantic communication system is proposed by exploiting knowledge graph. Moreover, a multi-scale compression network is designed for semantic compression to reduce data transmission volume while guaranteeing the detection performance. Furthermore, an object detection scheme is proposed by using the knowledge graph to overcome channel noise interference and compression distortion. Simulation results conducted on the practical aerial image dataset demonstrate that compared to the benchmark systems, our proposed system has superior detection accuracy, communication robustness and computation efficiency even under high compression rates and low signal-to-noise ratio (SNR) conditions.

\end{abstract}


\section{Introduction}
Due to the advantages of flexible deployment, low cost and diverse promising technique equipment, unmanned aerial vehicles (UAVs) are used for object detection in various civil and military applications, such as military reconnaissance, environmental monitoring, and urban planning\cite{srivastava2021survey}. However, since the battery capacity, computation and communication resources of UAV are finite, the existing UAV object detection systems are subject to severe challenges, such as limited endurance time and high computation latency, which constrains the wide application and detection performance. Thus, in order to overcome those challenges, it is imperative to develop new UAV-based object detection systems\cite{wu2022deep}.

\begin{figure*}[!t]
    \centering
    \includegraphics[width=5.5 in]{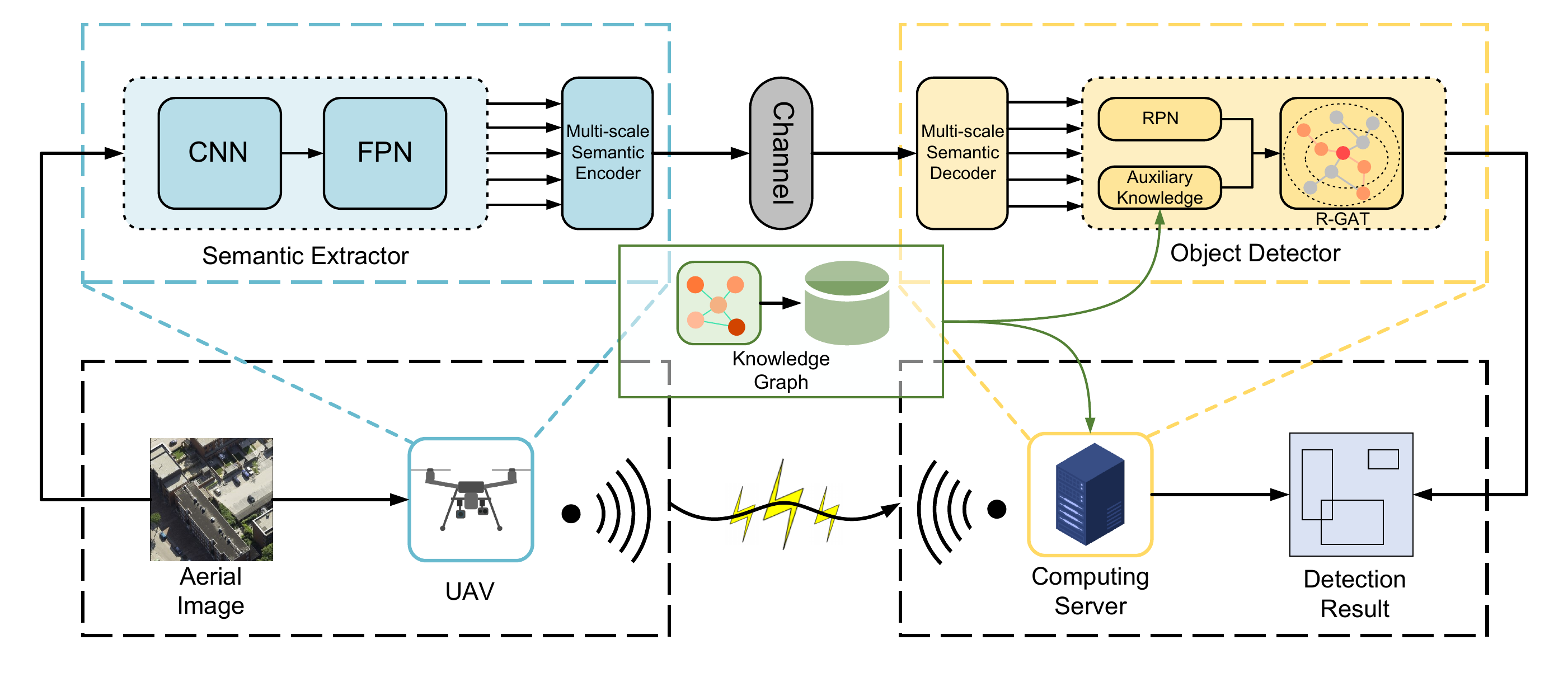}
    \caption{The proposed UAV cognitive semantic communication framework.}
\end{figure*}

The existing frameworks for UAV-based object detection can be mainly classified into two categories. One is to deploy the object detection algorithm on the UAV, utilizing the computation resources of the UAV for detection algorithms. The authors in\cite{deng2020energy} proposed an energy-efficient and real-time UAV-based object detection system on an embedded platform. The other one is to transmit the detection tasks to the ground server and the detection algorithms are performed in the ground\cite{wu2021unified}. In this system, UAV collects the image data and transmits them to the ground computing server over wireless communication networks, while the computing server performs the detection algorithms with powerful computation capability. In \cite{ozer2023offloading}, the authors designed an object detection system to transmit the deep learning (DL) powered vision tasks from the UAV to a remote server. Although the computation overhead of the UAV is reduced, the detection accuracy is still limited due to the bandwidth limitation and interference influence of wireless communication.

Fortunately, semantic communication is promising to overcome the above-mentioned issues, which can realize a good trade-off between the detection accuracy and the required energy, computation as well as communication resources. It was proposed by Weaver and Shannon \cite{xu2012opportunistic}. Unlike conventional wireless communication systems that prioritize the successful transmission of symbols between the transmitter and the receiver, semantic communication is to transmit the semantic information at the transmitter and interpret the semantic meaning at the receiver\cite{zhou2023cognitive}. Semantic communication can significantly reduce the data transmission volume while enhancing the system robustness and improving the detection performance\cite{qin2021semantic}.

The existing semantic communication can be categorized into three paradigms. The first paradigm aims to reconstruct the raw data at the receiver. The authors in \cite{bourtsoulatze2019deepjscc} proposed a DL-based joint source and channel coding (deepJSCC) technique for wireless image transmission. Based on deepJSCC, the authors in \cite{xu2021wireless} proposed a method called attention DL-based JSCC (ADJSCC) that can successfully operate with different signal-to-noise ratio (SNR) levels during transmission. The second paradigm is task-oriented semantic communication. This paradigm focuses on transmitting semantic features to enhance performance of the specific tasks. In \cite{xie2022task}, semantic communication was utilized for performing various tasks and demonstrated superior performance compared to the conventional methods. The last one is the cognitive semantic communication proposed in \cite{zhou2023cognitive}, which exploits knowledge graph to achieve semantic compression and semantic inference in order to improve the transmission efficiency and accuracy. In \cite{zhou2023cognitive}, the authors have demonstrated that the exploitation of knowledge graph can significantly improve the communication efficiency and reliability, and proposed two cognitive semantic communication frameworks for the single-user and multi-user semantic communication scenarios.

Motivated by the above-mentioned facts, in this paper, a UAV cognitive semantic communication system for object detection is proposed by exploiting knowledge graph. In order to reduce data transmission volume while ensuring detection accuracy, a multi-scale compression network is designed for semantic compression. Moreover, a detection scheme is proposed by exploiting auxiliary knowledge of the knowledge graph to overcome the channel noise interference and compression distortion. Simulation results conducted on the practical aerial image dataset demonstrate that our proposed system has superior detection accuracy, communication robustness and computation efficiency even under high compression rates and low SNR conditions, compared to the benchmark systems.

The remainder of this paper is organized as follows. Section II presents our proposed UAV cognitive semantic communication system. Section III elaborates on the details of our proposed framework for the implementation. Simulation results are provided in Section IV. Finally, the paper concludes with Section V.

\section{UAV Cognitive Semantic Communication For Object Detection}
In this section, a UAV cognitive semantic communication system is proposed, which consists of three modules, semantic extractor, multi-scale semantic encoder and decoder, and object detector. Semantic extractor and multi-scale semantic encoder are deployed on a computation-resource-constrained UAV, while the multi-scale semantic decoder and object detector are implemented on a computing server with sufficient computation resources. Moreover, our proposed UAV cognitive semantic communication system exploits a knowledge graph to enhance detection performance and communication robustness. Our proposed framework is shown in Fig. 1.

Inspired by the computer vision community, we exploit an innovative semantic extractor for the UAV semantic communication system to extract the semantic features of aerial images. In order to alleviate computation resources consumption, a light convolutional neural network (CNN) is used as the backbone of the semantic extractor. To tackle the challenge of multi-scale variation prevalent in aerial images, we incorporate the feature pyramid network (FPN) into the semantic extractor. 

Moreover, the purpose of semantic encoding and decoding is to reduce semantic redundancy while compressing semantic information. Compared to the traditional semantic communication systems, our semantic encoder and decoder adopt a novel parallel architecture. Specifically, five single-scale encoders and decoders, which correspond to the five scales of semantic features output by FPN, make up the parallel structure of the multi-scale semantic encoder and decoder. Through this design, the encoder and decoder are not only better equipped to handle visual semantic at different scales, but also tend to retain more useful semantic information and productively reduce redundancy. 

\begin{figure*}[!t]
    \centering
    \includegraphics[width=6 in]{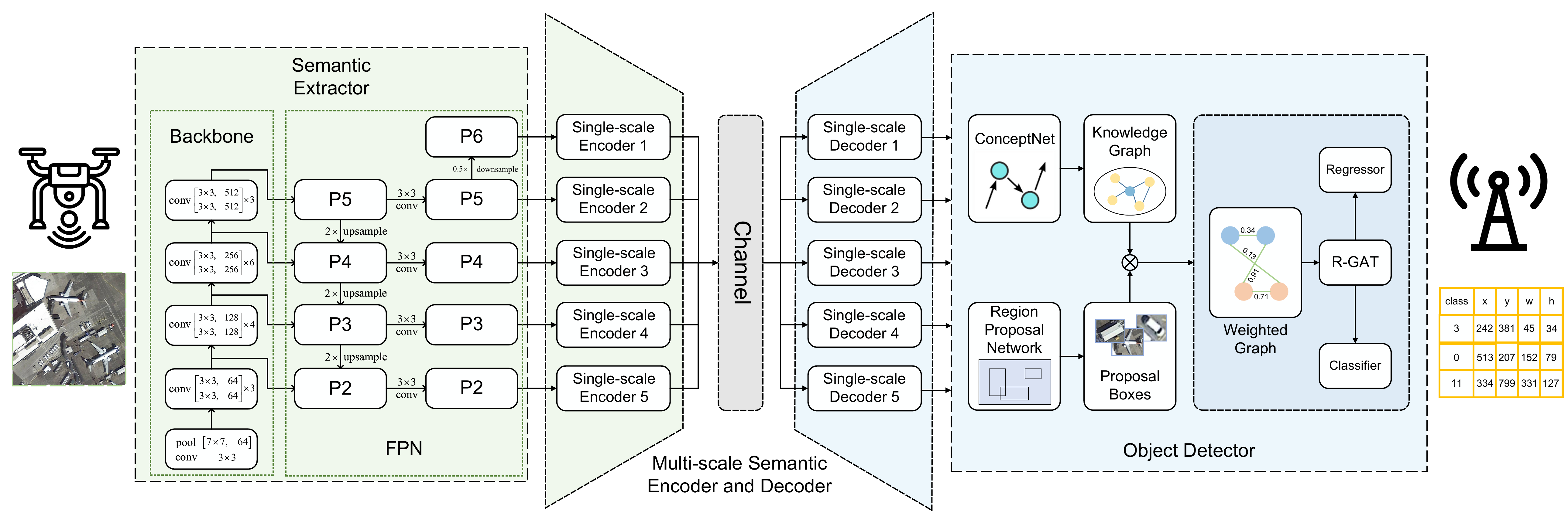}
    \caption{The network structure of the proposed UAV cognitive semantic communication system.}
\end{figure*}

After compression of the multi-scale semantic encoder, the semantic information is transmitted via wireless communication from the UAV to the computing server and is affected by channel noise. Additive white Gaussian noise (AWGN) channel and fading channel are considered in our work. The channel output $\hat{\boldsymbol{z}} \in \mathbb{C}^{k}$ can be expressed as
\begin{align}
    \hat{\boldsymbol{z}}=\boldsymbol{z}+\boldsymbol{\omega},
\end{align}
where the vector $\boldsymbol{z} \in \mathbb{C}^{k}$ is the channel input and the vector $\boldsymbol{\omega} \in \mathbb{C}^{k}$ consists of indpendent and identically distributed (i.i.d) samples with the distribution $\mathbb{CN}(0, \sigma^2I)$. $\sigma^2$ is the noise power and $\mathbb{CN}(\cdot,\cdot)$  denotes a circularly symmetric complex Gaussian distribution. The proposed system can be easily extended to other differentiable channels. In case of a fading channel, the process can be represented as
\begin{align}
    \hat{\boldsymbol{z}}=g\boldsymbol{z}+\boldsymbol{\omega},
\end{align}
where $g \in \mathbb{C}$ is the channel gain. On the computing server, the multi-scale semantic decoder reconstructs the noisy multi-scale semantic features.

Due to the inevitable noises, semantic expressed in the recovered messages may not match that of transmitter. Since the volume of semantic features is much less than source signals, few transmission mistakes result in severe semantic distortion. To overcome these issues, a pioneering detection scheme is proposed, which fuses the auxiliary knowledge of the knowledge graph and the visual semantic features into a weighted graph to achieve a unified global semantic understanding of the image, and dramatically improves the detection accuracy. As the cornerstone of cognitive semantic communication, the knowledge graph is a semantic network that represents relationships between entities in the form of graphs. Typically, it employs triples (head, relation, tail) or (entity, attribute, value) to convey factual information. 

\section{System Implementation}
In this section, the details for the implementation of the proposed system are presented. Fig. 2 shows the network structure of the proposed UAV cognitive semantic communication system.

\subsection{Semantic Computing Module}
The input of the system is the image captured by the UAV, denoted by $\mathbf{S}_I\in \mathbb{R}^{3 \times w \times h}$, where $w$ and $h$ are the width and height of the image, respectively. Then, the image is inputted into the semantic extractor to extract semantic features. Considering the limited computation resources of the UAV, ResNet-34 is used as the backbone network for the semantic extractor. Note that the ResNet-34 is pre-trained on ImageNet and boasts a robust capability to extract semantic information. For aerial images, it is challenging to detect objects in different scales, especially the small objects. To mitigate this issue, FPN is utilized as a semantic extractor component. In this case, the semantic extractor can obtain better-quality multi-scale semantic features for object detection. The process of extracting semantic features can be expressed as
\begin{align}
    \mathbf{P}_I = \mathcal{SE}_I(\mathbf{S}_I; \alpha_I),
\end{align}
where $\mathbf{P}_I=\{\mathbf{P}_2, \mathbf{P}_3, \mathbf{P}_4, \mathbf{P}_5, \mathbf{P}_6\}$ is the multi-scale semantic features, $\mathbf{P}_i \in \mathbb{R}^{256 \times s_i \times s_i}$ and $s_i$ is the size of semantic features; $\alpha_I$ is the trainable parameters.

\begin{figure}[t]
    \centering
    \includegraphics[width=3 in]{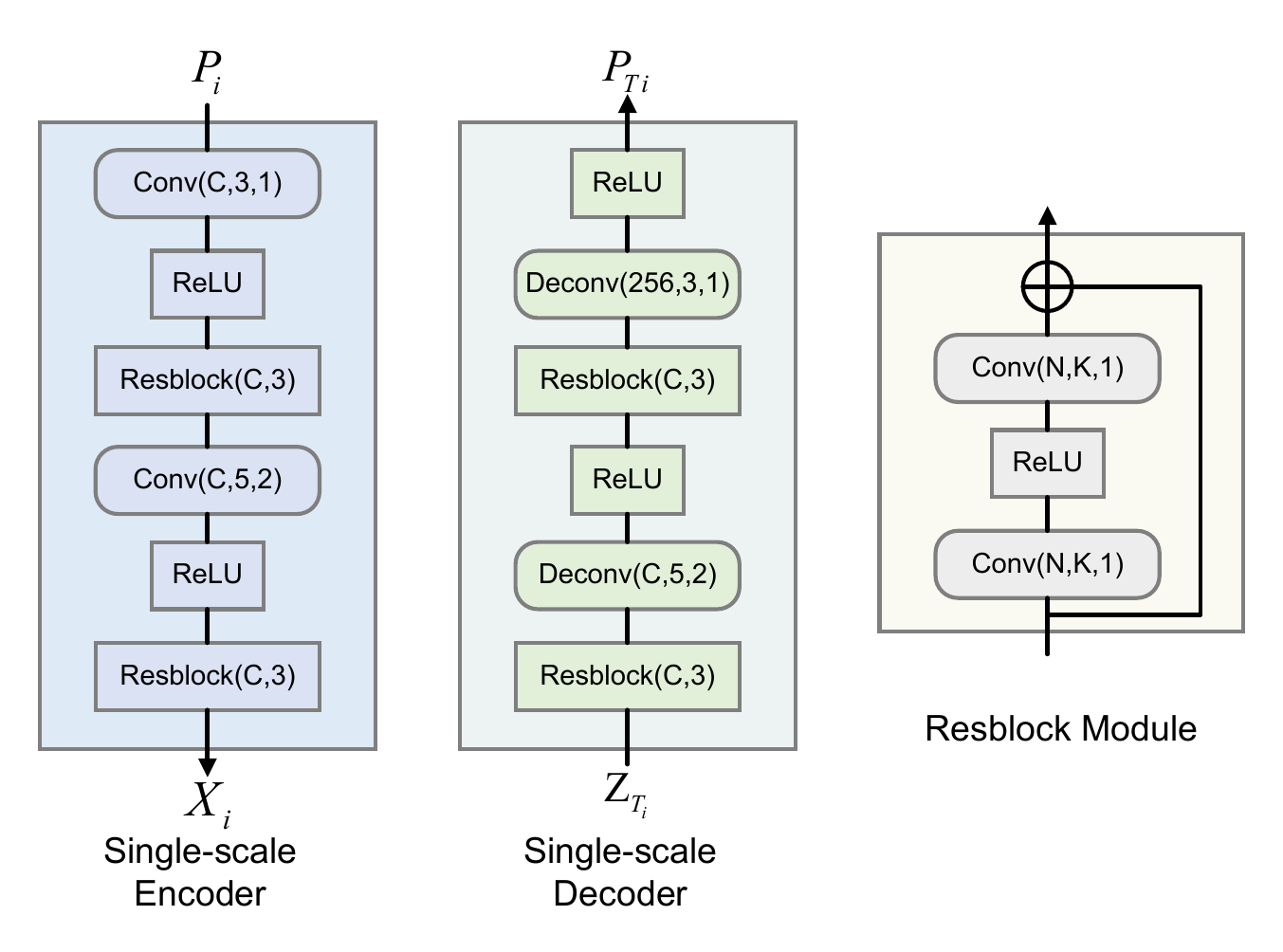}
    \caption{The details of the single scale encoder, decoder, and resblock. “Conv(N,K,S)” and “Deconv(N,K,S)” represent the convolution and deconvolution operations with the output channel, the kernel size, and the stride as $N$, $K \times K$ and $S$, respectively.}
\end{figure}

In order to reduce the number of the transmitted symbols, multi-scale semantic features are encoded as $\mathbf{X}_I=\{\mathbf{X}_2, \mathbf{X}_3, \mathbf{X}_4, \mathbf{X}_5, \mathbf{X}_6\}$ by using the multi-scale semantic encoder as
\begin{align}
    \mathbf{X}_I = \mathcal{ME}_I(\mathbf{P}_I; \beta_I),
\end{align}
where $\beta_I$ is the trainable parameters of the multi-scale semantic encoder. The multi-scale semantic encoder is a parallel residual compression module, which is designed to further compress the semantic features. The details of the single-scale encoder are shown in Fig. 3. After processing through the multi-scale semantic encoder, the dimension of semantic features is transformed as $\mathbf{X}_i \in \mathbb{R}^{C \times \frac{s_i}{2} \times \frac{s_i}{2}}$, where the number of channel $C$ is determined by the bandwidth compression ratio $R$. $R=k/n$, where $n$ is the size of image and $k$ is the input size of the channel. 
The encoded semantic features at each scale are reshaped as a vector and converted into the complex signal denoted as $\tilde{\boldsymbol{z}}_i \in \mathbb{C}^{k_i}$, $k_i$ is the length of the vector. After that, the vector is normalized as
\begin{align}
    \boldsymbol{z}_i=\sqrt{k_iP}\frac{\tilde{\boldsymbol{z}}_i}{\sqrt{\tilde{\boldsymbol{z}}_i^\ast\tilde{\boldsymbol{z}}_i}}, \label{eq3}
\end{align}
where $\tilde{\boldsymbol{z}}_i^\ast$ is the conjugate transpose of $\tilde{\boldsymbol{z}}_i$. The channel input $\boldsymbol{z}_i$ satisfies the average transmit power constraint $P$. Then, $\boldsymbol{z}_i$ is transmitted over a noisy channel, which can be represented as eq.(1) and eq.(2).

\subsection{Knowledge Graph Enhanced Object Detection}

At the computing server, the channel output $\hat{\boldsymbol{z}}_i$ is concatenated and reshaped to the original dimension like $\mathbf{X}_I$ before reconstructing the semantic information. The reshaped semantic information is denoted as $\mathbf{Z}_T$. After that, the multi-scale semantic decoder performs the reconstruction process, given as
\begin{align}
    \mathbf{P}_T = \mathcal{MD}_T(\mathbf{Z}_T; \beta_T),
\end{align}
where $\beta_T$ is the trainable parameters. The multi-scale semantic decoder shares a symmetrical structure with the multi-scale semantic encoder and uses deconvolutional layers to reconstruct the semantic features. The details of the single-scale decoder are demonstrated in Fig. 3.

The object detector is the core component of the proposed system, utilizing the received noisy semantic features to obtain the final detection results. Compared to the conventional object detection, the semantic communication-driven object detection system may exhibit slightly lower detection accuracy. It is attributed to compression distortion and noise interference, which can lead to misjudgments of objects by the object detector. Thus, we propose a novel object detection scheme enhanced by knowledge graph, which significantly alleviates this issue. The knowledge graph is rich in auxiliary knowledge and has the capability to establish relationships and similarities among multiple objects in the image, enhancing the global semantic understanding of images and reducing the probability of misjudgment. To achieve this, we constructed a weighted graph. The weighted graph encompasses two categories of nodes and three types of weighted edges, aiming to exploit global information. Note that the original information is preserved in the weighted graph, thus the accuracy of the results is only enhanced compared to that without introducing additional information.

\begin{figure}[t]
    \begin{center}
    \subfigure[The proposed detection scheme.]{
    \includegraphics[width=2.8 in]{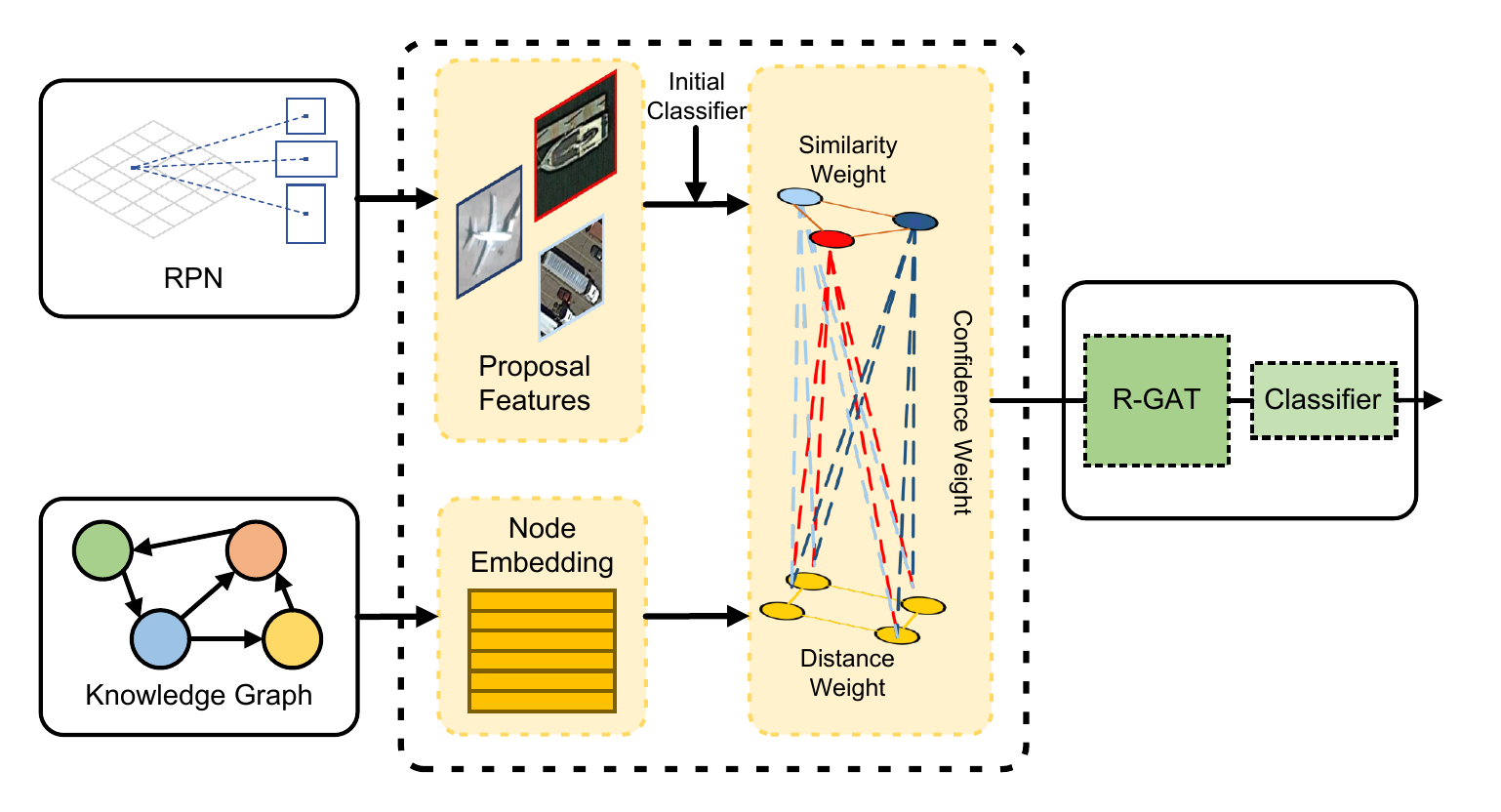}
    }
    \subfigure[An example of exploiting the knowsledge graph.]{
    \includegraphics[width=2.8 in]{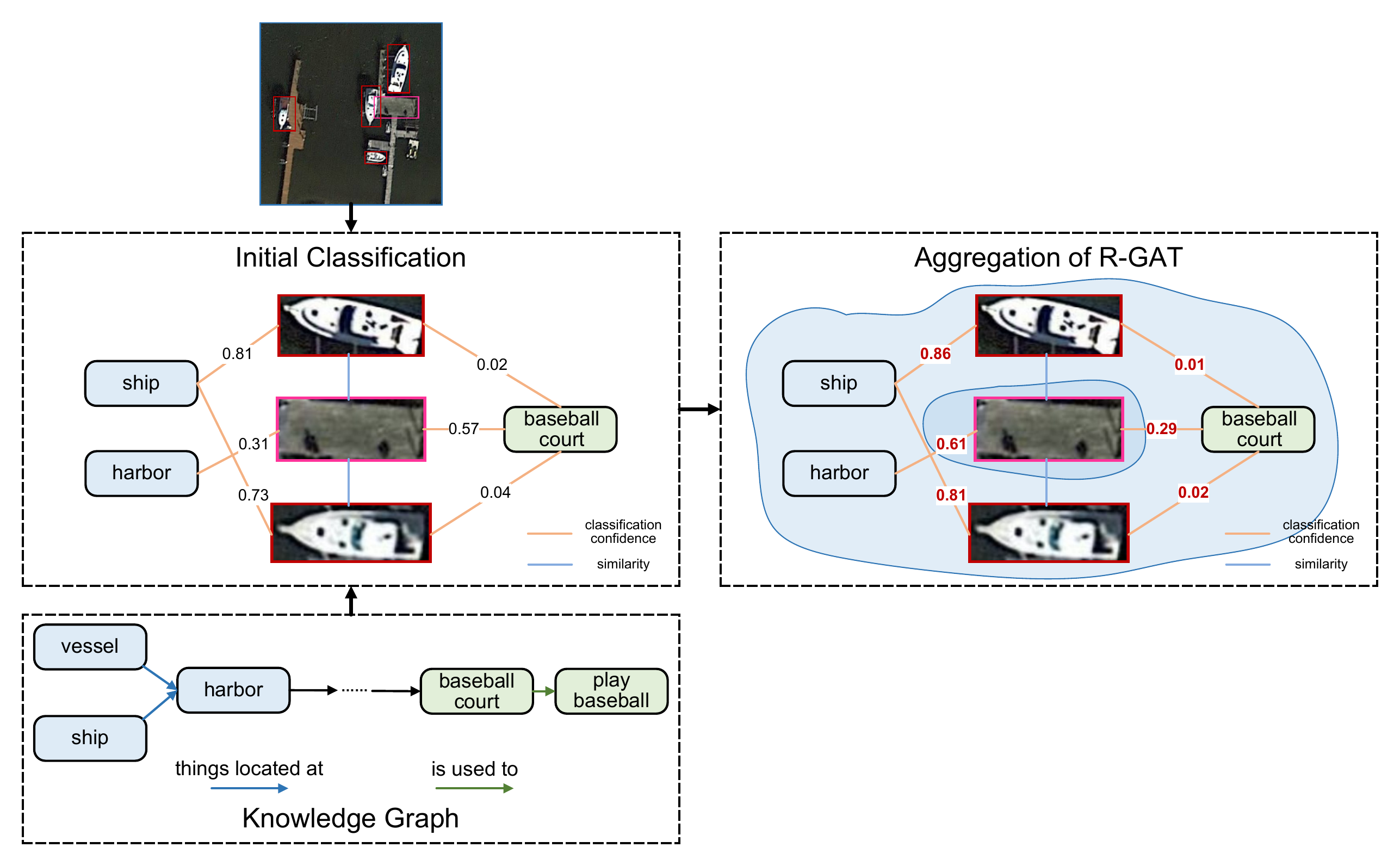}
    }
    \caption{The scheme of exploiting the auxiliary knowledge of the knowledge graph to improve model performance.}
    \end{center}
\end{figure}

The proposed detection scheme is shown in Fig. 4(a). The object detector is supplied with the reconstructed semantic features $\mathbf{P}_T$ to obtain the results. A region proposal network (RPN) and a knowledge graph comprise the object detector. The RPN is in a position to generate proposal boxes. Following that, the proposal boxes are fed into an initial classifier to obtain initial classification confidences, which are utilized to establish weights with nodes in the knowledge graph. Simultaneously, the weights between proposal boxes are derived from similarity function. The knowledge graph is responsible for introducing auxiliary knowledge to build relationships between objects in proposal boxes. To extract the auxiliary knowledge of the knowledge graph, the metapath2vec \cite{dong2017metapath2vec} model is adopted to embed the nodes of the knowledge graph. Then, the weights between nodes in the knowledge graph are derived through an embedding distance function. The embedding distance function and the similarity function are calculated by cosine similarity, which can be represented by 
\begin{equation}
    C_S(u, v) = \frac{u \cdot v}{\lVert u \rVert \lVert v \rVert},
\end{equation}
where $u$ and $v$ are the node embeddings, $\cdot$ is the inner product operateor.

After constructing the weighted graph, a relational graph attention network (R-GAT) aggregates global information to unify the global semantic information of an image. The enhanced proposal boxes are then placed into a classifier to provide final results. As illustrated in Fig. 4(b), it is challenging to differentiate the object within the pink bounding box based on the semantic features of the object in the initial classification, leading to misjudgment. By exploiting classification results of other objects in the same image and auxiliary knowledge of the knowledge graph, the model establishes a comprehensive global semantic understanding of the scene, thereby enabling correct classification.

\begin{figure*}[htp]
    \begin{center}
    \subfigure[R=1/6]{
    \includegraphics[width=2in]{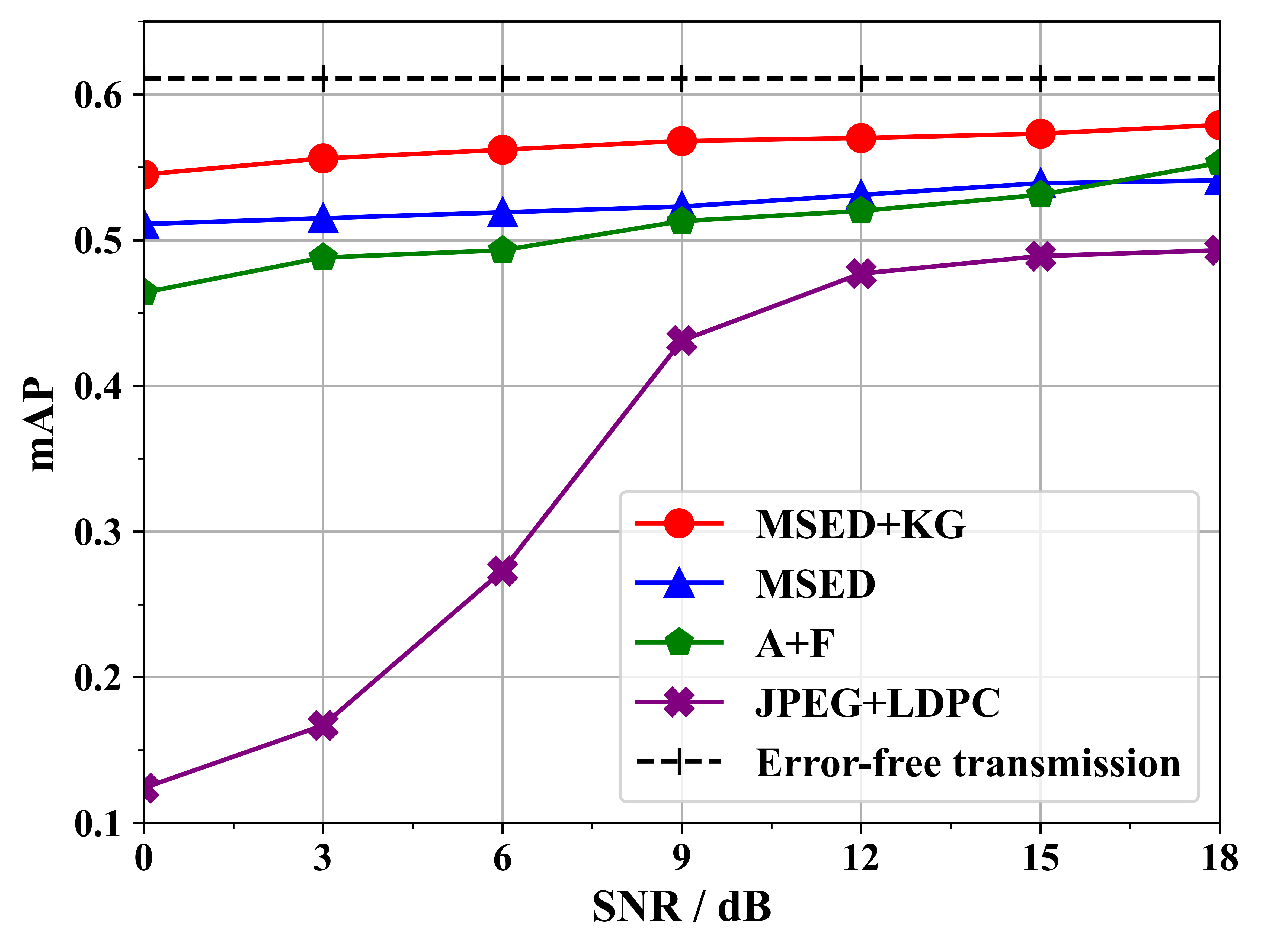}
    }\hspace{0.75in}
    \subfigure[R=1/12]{
    \includegraphics[width=2in]{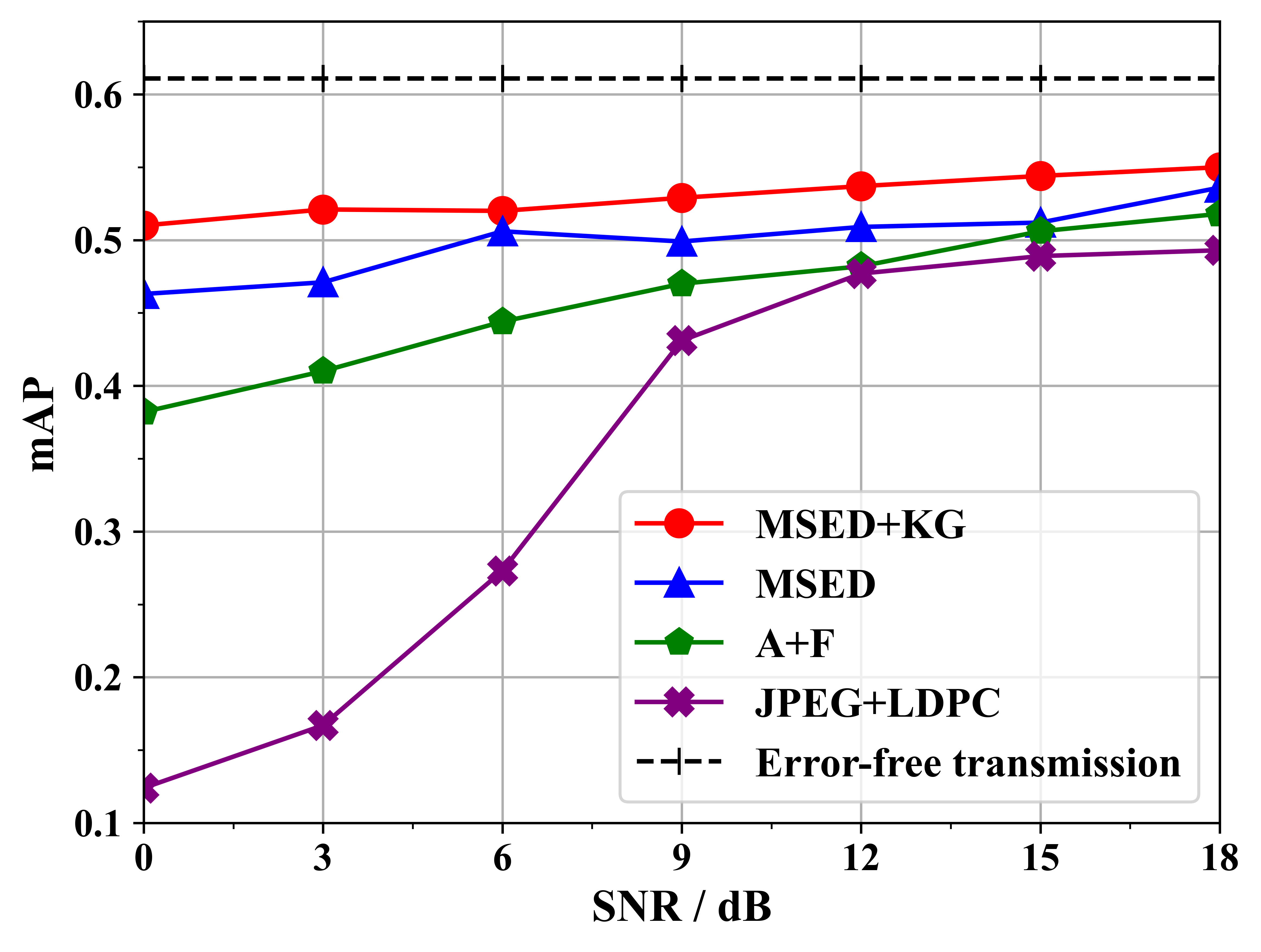}
    }
    \caption{mAP versus SNR under AWGN channel while R=1/6 and R=1/12.}
    \end{center}
\end{figure*}

\begin{figure*}[htp]
    \begin{center}
    \subfigure[R=1/6]{
    \includegraphics[width=2in]{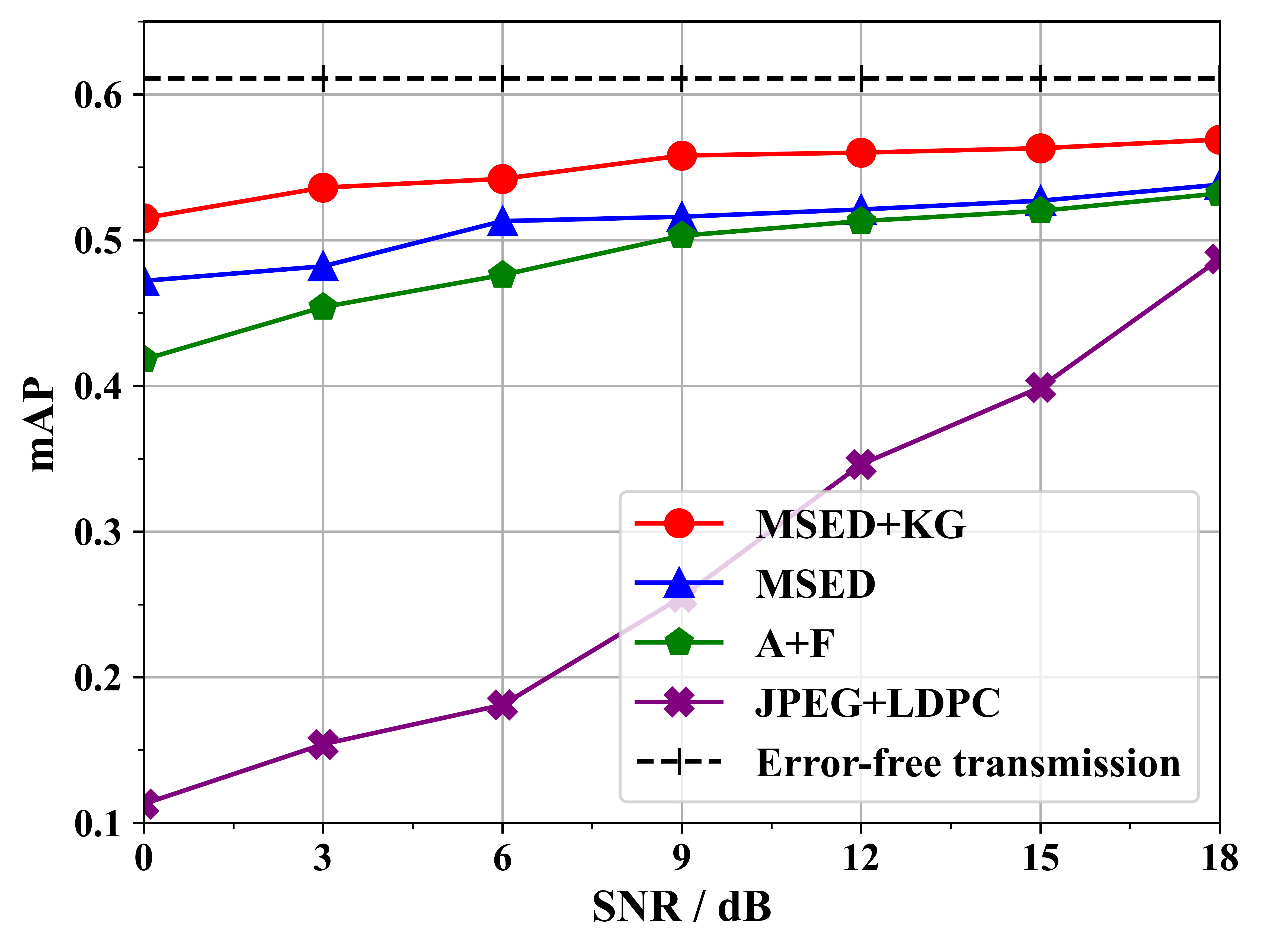}
    }\hspace{0.75in}
    \subfigure[R=1/12]{
    \includegraphics[width=2in]{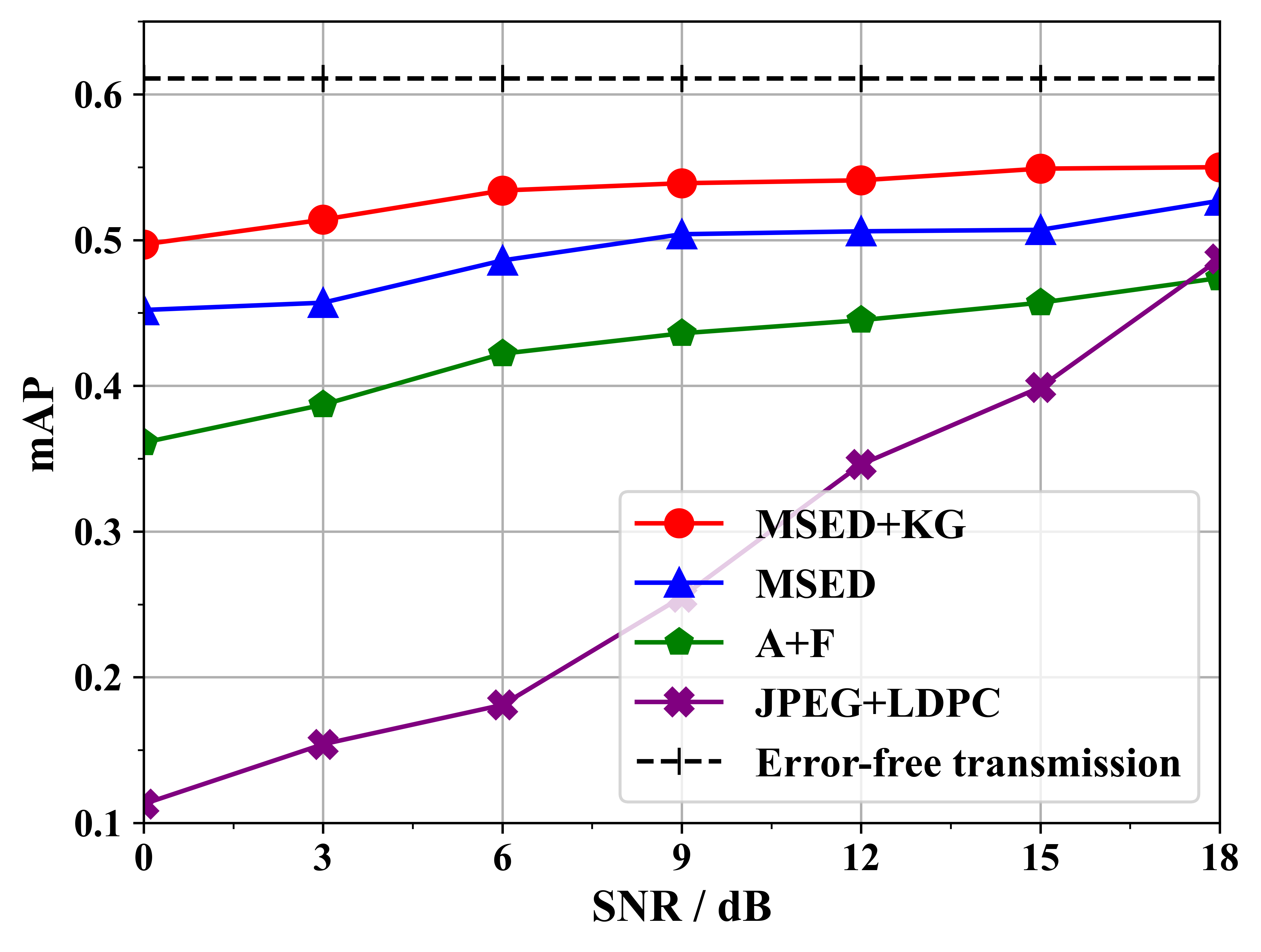}
    }
    \caption{mAP versus SNR under Rayleigh fading channel while R=1/6 and R=1/12.}
    \end{center}
\end{figure*}

\subsection{Model Learning}
To optimize the parameters of the proposed system, we minimize the loss function, which is defined as
\begin{align}
    L(\{p_i\}, \{t_i\})=\frac{1}{N_{cls}}\sum_{i}L_{cls}(p_i, p_i^\ast) \notag
    \\+ \lambda\frac{1}{N_{reg}}\sum_{i}p_i^\ast L_{reg}(t_i, t_i^\ast), 
\end{align}
where $i$ is the index of an anchor in a mini-batch, $p_i$ is the predicted probability that $i$ is an object. The ground truth label $p_i^\ast$ is 1 if the anchor is positive, and is 0 if the anchor is negative. $t_i$ is a vector representing the 4 parameterized coordinates of the predicted bounding box, and $t_i^\ast$ is that of the ground-truth box associated with a positive anchor. The classification loss $L_{cls}$ is log loss. The regression loss $L_{reg}(t_i, t_i^\ast)=R(t_i-t_i^\ast)$, where $R$ is smooth $L_1$ function. Using this loss function, all learnable parameters of the system are optimized.

\section{Simulation Results}
In this section, experiments are conducted to evaluate the efficacy of the proposed system across various communication channels, bandwidth compression ratios, and SNRs.

\subsection{Dataset and Baseline Systems}
The used dataset is the public large-scale dataset for object detection in aerial images (DOTA)\cite{xia2018dota}, which is a widely recognized benchmark object detection dataset in aerial scenes consisting of 15 categories. Each image is approximately $4000 \times 4000$ pixels in size and contains objects exhibiting a wide variety of scales.

Two classical systems are applied as comparison baselines. The first system utilizes ADJSCC for semantic communication, while the second is a conventional communication system that adopts JPEG with a 75\% compression rate for source coding, low-density parity-check codes (LDPC) with a coding rate of 1/3 for channel coding, and 64-QAM modulation. Faster RCNN with ResNet-50 is also employed for object detection in two baselines.

For our proposed system, RPN was applied to all scales of semantic features. Once the proposal boxes are generated, they are linked to the knowledge graph embedding as a weighted graph according to the proposed scheme. The embedding is then enhanced by R-GAT and inputted into two shared fully connected layers, serving as the input of the final classifier. For the R-GAT, we use a three-layer model. Adam optimizer with a learning rate of 10\textsuperscript{-4} optimizes the overall model.

\subsection{Experimental Results}

In all experimental results, we adopt the baseline ADJSCC+Faster RCNN denoted as A+F and the conventional communication system denoted as JPEG+LDPC. MSED+KG and MSED are the abbreviations for our proposed system with and without the knowledge graph. The average precision (mAP) metric is used to assess the detection performance of the system in experiments.

\begin{figure}[!t]
    \centering
    \includegraphics[width=1.9in]{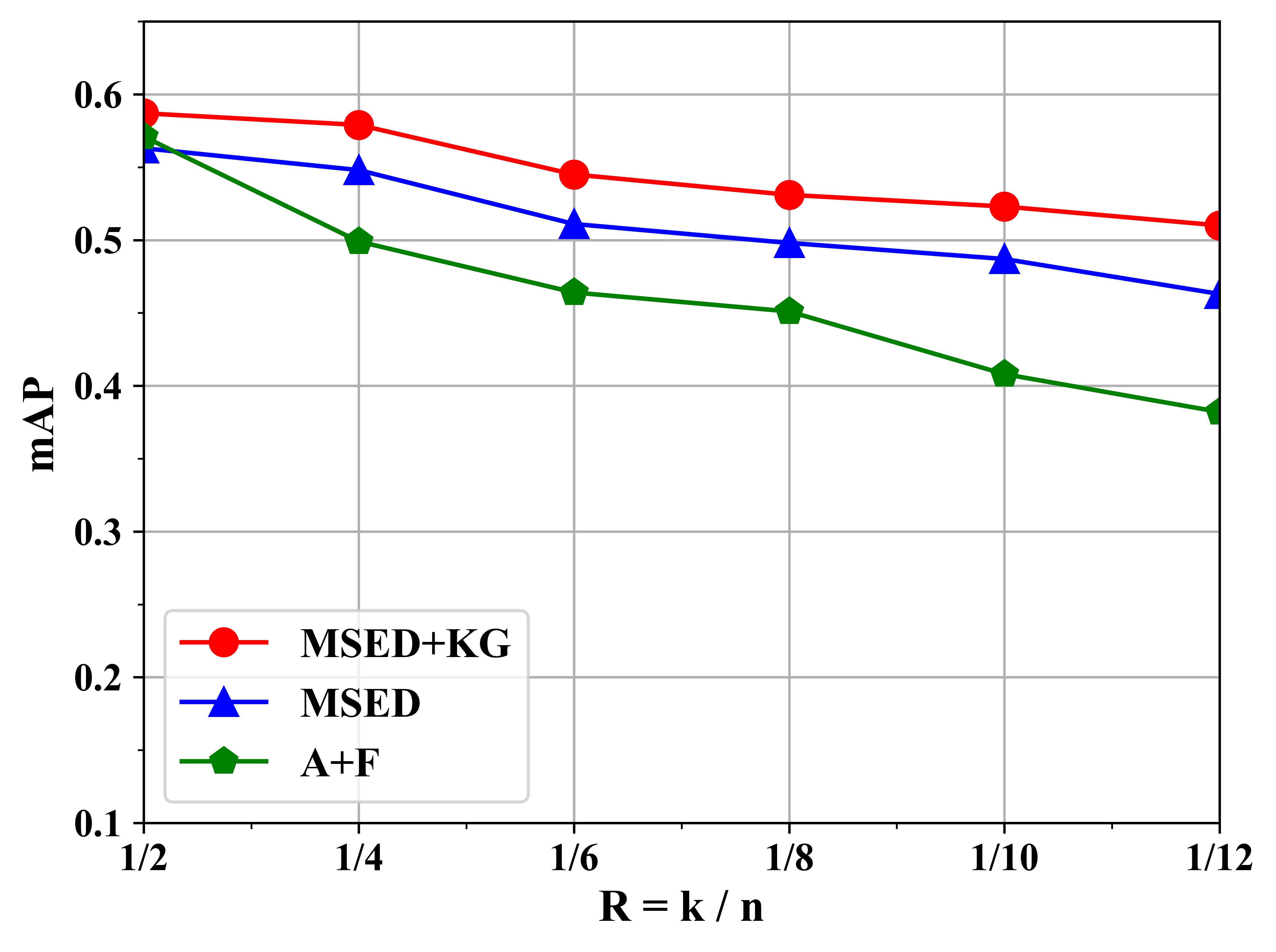}
    \caption{mAP versus bandwidth compression ratio R under AWGN channel while SNR=0.}
\end{figure}

\begin{table}[!t]
    \renewcommand{\arraystretch}{1.2}
    \centering
    \begin{tabular}{cccc}
    \hline
    \textbf{Method} & \textbf{No. Param} & \textbf{Additions} & \textbf{Multiplications} \\
    \hline
    MSED+KG & $5.1 \times 10^7$ & $3.9 \times 10^9$ & $3.8 \times 10^9$ \\
    A+F & $2.6 \times 10^8$ & $1.4 \times 10^{10}$ & $1.2 \times 10^{10}$ \\
    JPEG+LDPC & -- & $4.1\times 10^9$ & $4.3\times 10^9$ \\
    \hline
    \end{tabular}
    \caption{The number of the required parameters and the computational complexity of the models.}
\end{table}

Fig. 5 shows the mAP of all systems versus SNR under the AWGN channel when the bandwidth compression ratio R = 1/6 and R = 1/12. It is seen that MSED+KG consistently outperforms the MSED and the baseline, demonstrating superior accuracy, irrespective of the bandwidth compression ratio. This result indicates that the exploitation of auxiliary knowledge of the knowledge graph benefits for improving the accuracy of the system. Under high SNR conditions, MSED performs worse than A+F when R=1/6. However, this discrepancy does not occur when R=1/12. The reason is that the multi-scale semantic encoder and decoder conduct semantic compression and reconstruction, which has stronger tolerance to high compression rates compared to the data compression-restoration systems. 

To further verify the scalability and robustness of the proposed system, experiments are also conducted under Rayleigh channel. Fig. 6 presents the mAP of all systems versus SNR under the Rayleigh channel when the bandwidth compression ratio R = 1/6 and R = 1/12. Under these circumstances, with higher complexity, the mAP of all systems declines. However, compared to the proposed system, the baseline systems exhibit a more pronounced decrease in accuracy. The JPEG+LDPC and A+F experience an average accuracy drop of 7.4\% and 3.2\%, respectively, when R=1/12. In contrast, MSED+KG exhibits an average accuracy drop of 0.6\%. It is obvious that the proposed system is more competitive and robust in complex channel environments. 

Fig. 7 shows the detection accuracy of the proposed system and the baseline A+F at different bandwidth compression ratios under AWGN channel and SNR = 0. When the bandwidth compression ratio decreases, the proposed system is minimally affected, which further demonstrates its robustness. Furthermore, at the same mAP, the proposed system can achieve a higher bandwidth compression ratio. This benefit is particularly crucial in cases where bandwidth is constrained.

Table I compares the systems based on the number of required parameters and the computational complexity. For comprasion, the proposed system significantly decreases the computational complexity. In general, the proposed system can save computation resources and communication resources for UAV-based object detection.

\section{Conclusion}
A UAV cognitive semantic communication system was proposed for object detection by exploiting knowledge graph. Moreover, a multi-scale compression network was developed for semantic compression to reduce data transmission while ensuring the detection accuracy. Furthermore, a detection scheme was proposed by exploiting auxiliary knowledge of the knowledge graph to overcome channel noise interference and compression distortion. Simulation results demonstrated that the proposed system has superior detection accuracy, communication robustness and computation efficacy  under high compression rates and low SNR conditions.

\bibliographystyle{IEEEtran}
\bibliography{reference}

\begin{thebibliography}{10}
\providecommand{\url}[1]{#1}
\csname url@samestyle\endcsname
\providecommand{\newblock}{\relax}
\providecommand{\bibinfo}[2]{#2}
\providecommand{\BIBentrySTDinterwordspacing}{\spaceskip=0pt\relax}
\providecommand{\BIBentryALTinterwordstretchfactor}{4}
\providecommand{\BIBentryALTinterwordspacing}{\spaceskip=\fontdimen2\font plus
\BIBentryALTinterwordstretchfactor\fontdimen3\font minus
  \fontdimen4\font\relax}
\providecommand{\BIBforeignlanguage}[2]{{%
\expandafter\ifx\csname l@#1\endcsname\relax
\typeout{** WARNING: IEEEtran.bst: No hyphenation pattern has been}%
\typeout{** loaded for the language `#1'. Using the pattern for}%
\typeout{** the default language instead.}%
\else
\language=\csname l@#1\endcsname
\fi
#2}}
\providecommand{\BIBdecl}{\relax}
\BIBdecl

\bibitem{srivastava2021survey}
S.~Srivastava, S.~Narayan, and S.~Mittal, ``A survey of deep learning
  techniques for vehicle detection from uav images,'' \emph{J. Syst.
  Architect.}, vol. 117, p. 102152, 2021.

\bibitem{wu2022deep}
X.~Wu, W.~Li, D.~Hong, R.~Tao, and Q.~Du, ``Deep learning for unmanned aerial
  vehicle-based object detection and tracking: A survey,'' \emph{IEEE Geosci.
  Remote Sens. Mag.}, vol.~10, no.~1, pp. 91--124, 2022.

\bibitem{deng2020energy}
J.~Deng, Z.~Shi, and C.~Zhuo, ``Energy-efficient real-time uav object detection
  on embedded platforms,'' \emph{IEEE Trans. Comput.-Aided Design Integr.
  Circuits Syst.}, vol.~39, no.~10, pp. 3123--3127, 2020.

\bibitem{wu2021unified}
Q.~Wu, T.~Ruan, F.~Zhou, Y.~Huang, F.~Xu, S.~Zhao, Y.~Liu, and X.~Huang, ``A
  unified cognitive learning framework for adapting to dynamic environments and
  tasks,'' \emph{IEEE Wireless Commun.}, vol.~28, no.~6, pp. 208--216, 2021.

\bibitem{ozer2023offloading}
S.~Ozer, H.~E. Ilhan, M.~A. Ozkanoglu, and H.~A. Cirpan, ``Offloading deep
  learning powered vision tasks from uav to 5g edge server with denoising,''
  \emph{IEEE Trans. Veh. Technol.}, vol.~72, no.~6, pp. 8035--8048, 2023.

\bibitem{xu2012opportunistic}
Y.~Xu, J.~Wang, Q.~Wu, A.~Anpalagan, and Y.-D. Yao, ``Opportunistic spectrum
  access in cognitive radio networks: Global optimization using local
  interaction games,'' \emph{IEEE J. Sel. Topics Signal Process.}, vol.~6,
  no.~2, pp. 180--194, 2012.

\bibitem{zhou2023cognitive}
F.~Zhou, Y.~Li, M.~Xu, L.~Yuan, Q.~Wu, R.~Q. Hu, and N.~Al-Dhahir, ``Cognitive
  semantic communication systems driven by knowledge graph: Principle,
  implementation, and performance evaluation,'' \emph{IEEE Trans. Commun.}, pp.
  1--1, 2023.

\bibitem{qin2021semantic}
Z.~Qin, X.~Tao, J.~Lu, W.~Tong, and G.~Y. Li, ``Semantic communications:
  Principles and challenges,'' \emph{arXiv preprint arXiv:2201.01389}, 2021.

\bibitem{bourtsoulatze2019deepjscc}
E.~Bourtsoulatze, D.~Burth~Kurka, and D.~Gündüz, ``Deep joint source-channel
  coding for wireless image transmission,'' \emph{IEEE Trans. on Cogn. Commun.
  Netw.}, vol.~5, no.~3, pp. 567--579, 2019.

\bibitem{xu2021wireless}
J.~Xu, B.~Ai, W.~Chen, A.~Yang, P.~Sun, and M.~Rodrigues, ``Wireless image
  transmission using deep source channel coding with attention modules,''
  \emph{IEEE Trans. Circuits Syst. Video Technol.}, vol.~32, no.~4, pp.
  2315--2328, 2021.

\bibitem{xie2022task}
H.~Xie, Z.~Qin, X.~Tao, and K.~B. Letaief, ``Task-oriented multi-user semantic
  communications,'' \emph{IEEE J. Sel. Areas Commun.}, vol.~40, no.~9, pp.
  2584--2597, 2022.

\bibitem{dong2017metapath2vec}
Y.~Dong, N.~V. Chawla, and A.~Swami, ``metapath2vec: Scalable representation
  learning for heterogeneous networks,'' in \emph{Proceedings of the 23rd ACM
  SIGKDD international conference on knowledge discovery and data mining},
  2017, pp. 135--144.

\bibitem{xia2018dota}
G.-S. Xia, X.~Bai, J.~Ding, Z.~Zhu, S.~Belongie, J.~Luo, M.~Datcu, M.~Pelillo,
  and L.~Zhang, ``Dota: A large-scale dataset for object detection in aerial
  images,'' in \emph{Proceedings of the IEEE conference on computer vision and
  pattern recognition}, 2018, pp. 3974--3983.

\end{thebibliography}

\end{document}